\documentclass[aps,prl,amsmath,amssymb,floatfix,twocolumn,amsmath,superscriptaddress,twocolumn,nofootinbib,tighten,letterpaper]{revtex4-2}
\usepackage[colorlinks,linkcolor=blue,citecolor=blue,urlcolor=blue]{hyperref}
\usepackage{multirow}
\usepackage{subfigure}
\usepackage{color}
\usepackage{mathrsfs}
\usepackage{hyperref}
\usepackage[normalem]{ulem}
\usepackage{bm}

\newcommand{\bra}[1]{\left< #1 \right|}
\newcommand{\ket}[1]{\left| #1 \right>}

\usepackage{amssymb}
\usepackage{amsmath}
\renewcommand\vec[1]{\ensuremath\boldsymbol{#1}}

\usepackage{tabularray}

\usepackage{array} 
\newcolumntype{P}[1]{>{\centering\arraybackslash}p{#1}}

\definecolor{RowColor}{rgb}{0.88,1,0.9}

\usepackage{amsfonts, relsize, color, mathtools, physics}
\usepackage{graphics}
\usepackage{graphicx}
\usepackage{subfigure}
\usepackage{color}
\usepackage{comment}

\begin{document}
\title{Dirac fermions in non-Hermitian magnetic fields: Zero modes and index theorem}

\author{Christopher A. Leong}
\affiliation{Department of Physics, Lehigh University, Bethlehem, Pennsylvania 18015, USA}

\author{Bitan Roy}
\affiliation{Department of Physics, Lehigh University, Bethlehem, Pennsylvania 18015, USA}

\date{\today}

\begin{abstract}
In a Lorentz symmetric non-Hermitian (NH) Dirac theory, containing the canonical relativistic Hamiltonian accompanied by a masslike anti-Hermitian Dirac operator, when the associated NH parameter becomes spatially modulated it couples massless Dirac fermions as NH gauge fields. With specific choices of such resulting NH gauge potential, the system experiences NH magnetic fields. When a planar Dirac system encloses a finite flux of such NH magnetic fields, a manifold of spatially localized \emph{right or left} zero-energy eigenmodes appear in the spectrum, which we numerically anchor from microscopic realizations of NH magnetic fields on graphene's honeycomb lattice. Potential experimental platforms to test these predictions are discussed. Altogether, zero-energy NH flat bands of right or left modes promise fascinating future realizations of NH magnetic catalysis, strongly-coupled NH fractional topological phases, and NH chiral anomaly, to name a few.    
\end{abstract}

\maketitle

\emph{Introduction}.~Massless Dirac fermions subject to strong external magnetic fields support a topologically protected degenerate manifold of zero-energy states (also known as the zeroth Landau level), guaranteed by the Aharonov-Casher index theorem (ACIT)~\cite{aharonovcasher:1}. The number of localized modes within such a manifold is exactly equal to the magnetic flux quanta enclosed by the systems, irrespective of the spatial profile of the magnetic field, which can also be tested from lattice-based numerical simulations~\cite{aharonovcasher:2} and possibly also extends to curved hyperbolic spaces~\cite{aharonovcasher:3}. When Dirac fermions are confined to a flat plane, such an outcome is responsible for the unique sequence of quantum Hall plateaus, experimentally observed in graphene that can exclusively be attributed to the quasi-relativistic nature of its low-energy emergent quasiparticles~\cite{grapheneQHE:1, grapheneQHE:2}. In three dimensions, the manifold of zero-energy bound states become dispersive in the direction of the applied magnetic field, responsible for the celebrated chiral anomaly~\cite{chiralanomalyTh:1, chiralanomalyTh:2, chiralanomalyTh:3} that has recently been observed in quantum Weyl and Dirac crystals~\cite{chiralanomalyExp:Review}. The ACIT also applies to time-reversal-symmetric axial magnetic fields~\cite{axialMFTh:1, axialMFTh:2, axialMFTh:3}, whose associated axial zeroth Landau levels have been observed experimentally~\cite{axialMFEp:1, axialMFEp:2, axialMFEp:3, axialMFEp:4}. It relies on the relativistic character of quasiparticles, described by a Lorentz-invariant Dirac Hamiltonian, and on their minimal coupling to gauge fields.

\begin{figure}[t!]
\includegraphics[width=0.90\linewidth]{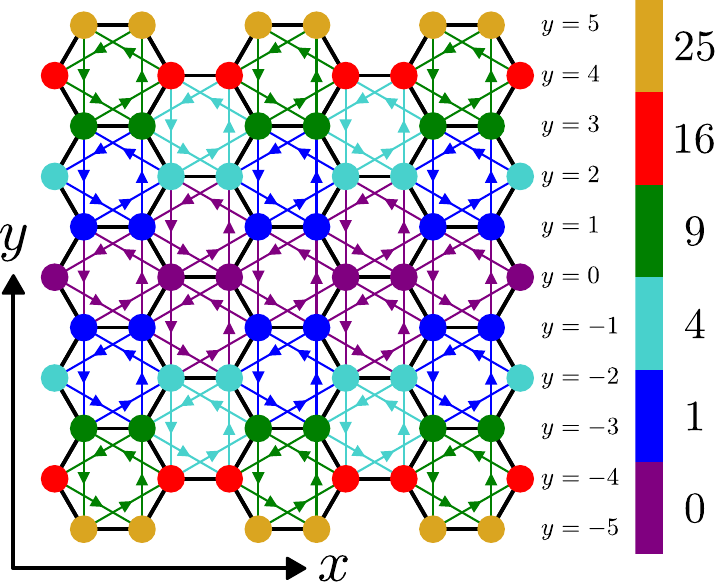}
\caption{Realizations of uniform non-Hermitian magnetic fields (NHMFs) on a honeycomb lattice with periodic (open) boundary conditions along the horizontal or $x$ (vertical or $y$) direction. Colored circles on each site denote the value of $\chi(\vec{r}) \sim y^2$, yielding a \emph{uniform} type-I NHMF. To generate a \emph{uniform} type-II NHMF, $\chi(\vec{r})$ results from purely imaginary next-nearest-neighbor (NNN) hopping amplitudes in the direction of arrows whose strength is determined by the smallest value of $y$ bordering each plaquette with $\chi(\vec{r}) \sim y^2$. Notice that the net flux through each plaquette is nonzero, since the currents within the sites of A and B sublattices of the honeycomb lattice flow in the same direction. For clarity, NNN currents across the system due to the periodic boundary condition in the $x$ direction are not shown explicitly.     
}~\label{fig:1_modeldiagram}
\end{figure}

\begin{figure}[t!]
\includegraphics[width=0.90\linewidth]{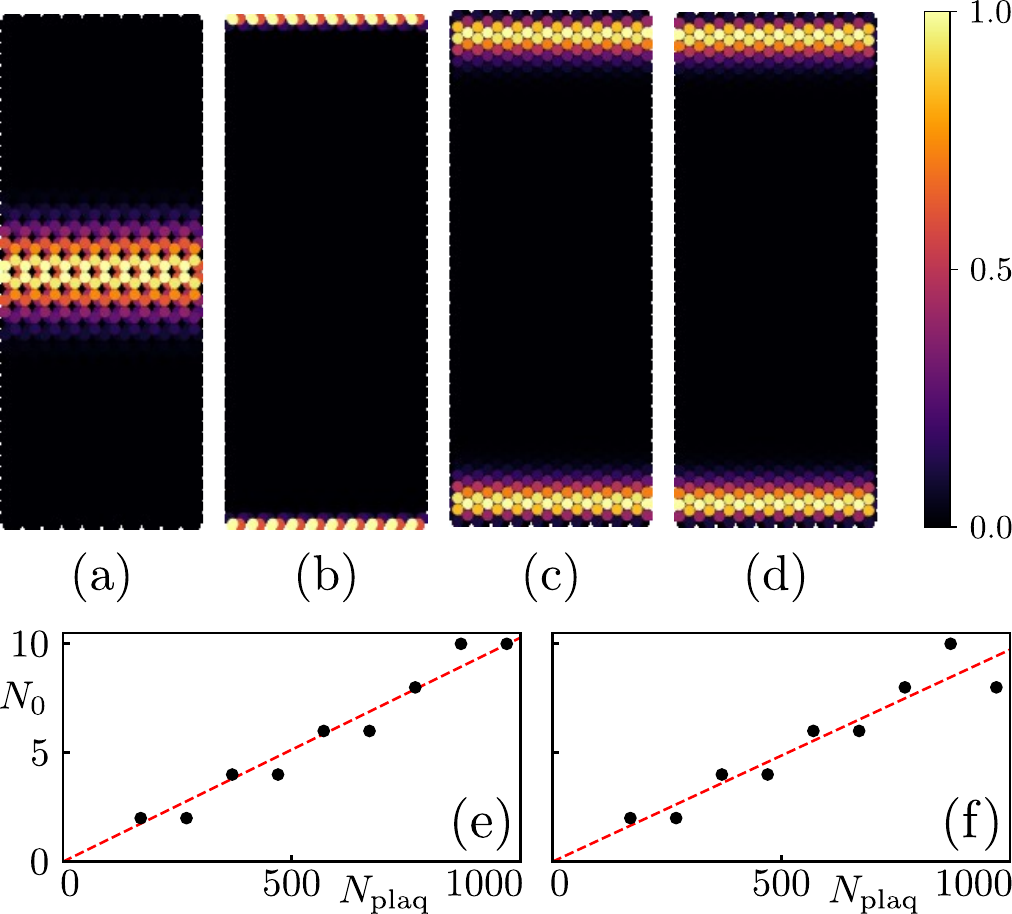}
\caption{Local density of states (normalized by its maximum value) on a honeycomb lattice, featuring massless Dirac fermions in the pristine condition, with periodic (open) boundary conditions along the horizontal or $x$ (vertical or $y$) direction, containing a total of 89 layers (in the $y$ direction) and 3780 sites, computed from the two closest to zero-energy modes for their (a) and (c) left and (b) and (d) right eigenvectors. The results are shown in the presence of uniform (a) and (b) type-I and (c) and (d) type-II non-Hermitian magnetic fields (NHMFs) with $\beta=0.01$ and $\beta=0.001$, respectively. With a type-I NHMF, left (right) zero-energy eigenmodes are normalizable (non-normalizable) and live in the bulk (on the $y$-directional edges) of the system. When the direction of type-I NHMF is flipped, the localization properties of the left and right zero-energy eigenmodes get reversed (not shown explicitly). By contrast, a type-II NHMF supports both normalizable and non-normalizable left and right zero-energy eigenmodes, and we observe a superposition of these two types of modes that are peaked closer to the boundary, but not on the edges of the system in the $y$ direction, compare with (b). When the direction of type-II NHMF is reversed, the situation remains unchanged. Scaling of the number of near zero-energy modes $N_0$, computed within the energy window $(-0.125,0.125)$, with the number of plaquettes $N_{\rm plaq}$ in the system (proportional to the flux enclosed) in the presence of uniform (e) type-I ($\beta=0.01$) and (f) type-II ($\beta=0.001$) NHMF, showing almost \emph{linear} scaling (red lines).        
}~\label{fig:2_ldos}
\end{figure}

We here show that retaining such quintessential features, it is conceivable to couple a collection of gapless Dirac fermions to masslike non-Hermitian magnetic fields (NHMFs) that in turn bind localized zero energy modes of, however, right or left eigenmodes. The outcomes can also be verified from honeycomb lattice-based microscopic realizations of NHMFs (Fig.~\ref{fig:1_modeldiagram}) and the spatial profiles of the resulting zero-energy modes (Fig.~\ref{fig:2_ldos}). The construction is based on the following general principle.

\emph{Construction}.~The Dirac theory is manifestly Lorentz invariant, permitting two Lorentz scalars, namely (a) the Dirac kinetic Hamiltonian and (b) Dirac mass~\cite{Dirac:1, Dirac:2}. In $d$ spatial dimensions the former operator reads as $H_{\rm Dir}(\vec{k})= v \left( \Gamma_j k_j \right)$, where $j=1, \cdots, d$ and a summation over repeated indices is assumed throughout. Here, $v$ ($\ll c$ in quantum materials) is the Fermi velocity of quasi-relativistic fermions, playing the role of the speed of light ($c$), $\Gamma_j$ are Hermitian matrices satisfying the anticommuting Clifford algebra $\{ \Gamma_j, \Gamma_k \} = 2 \delta_{jk}$ with $\delta_{jk}$ as the Kronecker delta symbol, and $k_j$ are spatial components of momentum ($\vec{k}$). A Dirac mass is represented by a Hermitian matrix $M$, satisfying $\{ M, \Gamma_j  \}=0$ and $M^2={\mathbf I}$ (identity matrix). Dimensionality and explicit representation of the $\Gamma$ matrices and $M$ depend on $d$ and microscopic details, which we discuss later. In terms of these two Hermitian Lorentz scalars, we introduce the Lorentz invariant non-Hermitian (NH) Dirac operator 
\begin{equation}~\label{eq:NHOperator}
H_{\rm Dir}^{\rm NH} (\vec{k})=H_{\rm Dir} (\vec{k}) + \alpha M H_{\rm Dir} (\vec{k}),
\end{equation}
where the real parameter $\alpha$ determines the strength of non-Hermiticity. The anti-Hermitian operator, second term in Eq.~\eqref{eq:NHOperator}, is called \emph{masslike} since it fully anticommutes with $H_{\rm Dir} (\vec{k})$, but vanishes when $\vec{k} \to 0$~\cite{NHDirac:1, NHDirac:2}.

To minimally couple Dirac fermions to masslike NH gauge fields within the framework of the above Lorentz invariant construction of NH Dirac theory, we take $\vec{k} \to \vec{A}(\vec{r})$ in its anti-Hermitian term, which can be considered as a NH vector potential with $\alpha$ now playing the role of the corresponding \emph{charge}. The NH Dirac operator is now  
\begin{equation}~\label{eq:NHGaugeField}
H_{\rm Dir}^{\rm NH} (\vec{A}, \vec{k})= v \Gamma_j \left( k_j + \alpha M A_j \right). 
\end{equation}
For convenience, we take $\alpha \vec{A} \to \vec{A}$ and set $v=1$. All the conclusions related to the bound states at zero energy are obtained starting from the above NH operator in $d=2$.

\emph{Zero modes}.~To showcase the existence of zero-energy bound states in the spectrum of $H_{\rm Dir}^{\rm NH} (\vec{A}, \vec{k})$ in $d=2$, it is sufficient to consider a minimal two-component Dirac system for which $\Gamma_1=\sigma_1$, $\Gamma_2=\sigma_2$, and $M=\sigma_3$, where $\{ \sigma_\mu \}$ is the set of Pauli matrices. Here, we work with the Coulomb or transverse gauge in which ${\boldsymbol \nabla} \cdot \vec{A}=0$. Then the NH vector potential can be parameterized as $\vec{A}(\vec{r})=(\partial_y \chi, -\partial_x \chi)(\vec{r})$ in terms of an arbitrary scalar function $\chi(\vec{r})$ such that the NHMF reads as $\vec{B}(\vec{r})={\boldsymbol \nabla} \times \vec{A}=-\nabla^2 \chi(\vec{r}) \hat{z}$, where $\hat{z}$ is the unit vector in the $z$ direction. Then $H_{\rm Dir}^{\rm NH} (\vec{A},\vec{k})$ takes the form 
\begin{eqnarray}~\label{eq:NHMagSimilarity}
H_{\rm Dir}^{\rm NH} (\vec{A},\vec{k}) \equiv H_{\rm Dir}^{\rm NH} (\chi,\vec{k})= e^{-\chi(\vec{r})} \; H_{\rm Dir}(0,\vec{k}) \; e^{\chi(\vec{r})}, 
\end{eqnarray}
which is related to the original Dirac Hamiltonian without any gauge potential through a space-dependent \emph{similarity} transformation, sourcing NHMFs.

We set out to find localized zero-energy left and right eigenmodes in the spectrum of $H_{\rm Dir}^{\rm NH} (\vec{A},\vec{k})$, obtained from the solutions of the following equations
\begin{eqnarray}~\label{eq:secularzeromodes}
\bra{\Psi^{\rm L}_0} H_{\rm Dir}^{\rm NH} (\chi,-i {\boldsymbol \nabla})=0
\; \text{and} \;
H_{\rm Dir}^{\rm NH} (\chi, -i {\boldsymbol \nabla}) \ket{\Psi^{\rm R}_0}=0,
\end{eqnarray}
respectively. From Eq.~\eqref{eq:NHMagSimilarity}, it can then be shown that
\begin{equation}~\label{eq:zeromodes}
\bra{\Psi^{\rm L}_0} \equiv \bra{\Psi^{\rm L}_0 (0)} e^{\chi(\vec{r})} 
\; \text{and} \;
\ket{\Psi^{\rm R}_0} \equiv e^{-\chi(\vec{r})} \ket{\Psi^{\rm R}_0 (0)},  
\end{equation}
where $\bra{\Psi^{\rm L}_0 (0)}$ and $\ket{\Psi^{\rm R}_0 (0)}$, are respectively the left and right zero-energy eigenvectors of $H_{\rm Dir}(0,-i {\boldsymbol \nabla})$, given by  
\begin{align}
\bra{\Psi^{\rm L}_0 (0)} &= r^{j} e^{i \; (j \phi)} \left( 1 \quad 0 \right) 
\; \text{or} \;
r^{j} e^{-i \; (j \phi)} \left( 1 \quad 0 \right) \nonumber  \\
\text{and} \:
\ket{\Psi^{\rm R}_0 (0)} &= r^{j} e^{i \; (j \phi)} \left( \begin{array}{c} 1 \\ 0 \end{array} \right) 
\; \text{or} \; 
r^{j} e^{-i \; (j \phi)} \left( \begin{array}{c} 0 \\ 1 \end{array} \right) 
\end{align}
with $\phi$ as the azimuthal angle, $j$ as an integer, $r=|\vec{r}|$, and $i=\sqrt{-1}$. Notice that $\bra{\Psi^{\rm L}_0 (0)}$ and $\ket{\Psi^{\rm R}_0 (0)}$ are eigenstates of $\sigma_3$ with eigenvalues $\pm 1$, which results from the unitary particle-hole or sublattice symmetry of the Dirac Hamiltonian, given by $\{ H_{\rm Dir}^{\rm NH} (\chi, -i {\boldsymbol \nabla}), \sigma_3\}=0$ for any $\chi(\vec{r})$. Neither $\bra{\Psi^{\rm L}_0 (0)}$ nor $\ket{\Psi^{\rm R}_0 (0)}$ is individually normalizable and their biorthogonal product is also not normalizable. Therefore, between zero mode solutions from Eq.~\eqref{eq:zeromodes}, either $\bra{\Psi^{\rm L}_0}$ or $\ket{\Psi^{\rm R}_0}$ is individually normalizable, while their biorthogonal product is not normalizable. For $\chi(r)>0$ ($\chi(r)<0$) we find normalizable right (left) zero-energy eigenmodes. A change in the sign of $\chi(\vec{r})$ flips the direction of the NHMF. Hence, nodal Dirac fermions subject to masslike NHMFs accommodate a manifold of normalizable right or left zero-energy modes. The number of such self normalizable zero modes, setting the degeneracy of the associated manifold, is found in the following way.

In terms of the Green's function of the Laplacian in two dimensions, the scalar function $\chi(\vec{r})$ can be related to the NHMF according to~\cite{aharonovcasher:1} 
\begin{equation}
\chi(\vec{r})= -\int \frac{d^2 \vec{r}^\prime}{2 \pi} \; \ln|\vec{r}^\prime - \vec{r}| \; B(\vec{r}^\prime)
\end{equation}
since $B(\vec{r})=-\nabla^2 \chi(\vec{r})$. Therefore, in the $|\vec{r}| \to \infty$ limit 
\begin{align}
\exp[-\chi(\vec{r})] \to \exp\left[\ln|\vec{r}| \int \frac{d^2 \vec{r}^\prime}{2 \pi} B(\vec{r}^\prime) \right]
=  r^{-\Phi/2 \pi},
\end{align}
where $\Phi =\int d^2\vec{r}^\prime B(\vec{r}^\prime)$ is the total flux of NHMF enclosed by the system. For $\alpha=\hbar=1$, the corresponding flux quanta $\Phi^{\rm NH}_0=2 \pi$, and $\Phi/2\pi \equiv \Phi/\Phi^{\rm NH}_0$ is the total NH magnetic flux quanta ($N$) enclosed by the system. Then the \emph{self-normalization} condition for the right zero-energy eigenmode, given by  
\begin{equation}~\label{eq:selfnormalization}
\int \frac{d^2 \vec{r}}{(2\pi)^2} \; \bra{\Psi^{\rm R}_0} \Psi^{\rm R}_0 \rangle =1 
\end{equation}
can be satisfied only for the integer values of $j=0,1, \cdots, N-1$. Therefore, when a planar gapless relativistic system encloses $N$ number of flux quanta of masslike NHMFs, the system supports $N$ number of self-normalizable right zero-energy eigenmodes, irrespective of the profile of NHMFs.  

\emph{Lattice model}.~Next we set out to anchor the predictions obtained from the continuum model for massless Dirac fermions, coupled to masslike NHMFs, from a concrete lattice-regularized model. We consider graphene's honeycomb lattice, a canonical microscopic setup fostering massless Dirac fermions as low-energy emergent quasiparticle excitations near the two inequivalent corners of the hexagonal Brillouin zone located at $\pm {\bf K}$, where ${\bf K}=(1/2, 1/2\sqrt{3}) 4\pi/\sqrt{3}a$, where $a$ is the lattice spacing. The corresponding Dirac Hamiltonian is obtained by expanding the Bloch Hamiltonian associated with the nearest-neighbor tight-binding model for spinless fermions with hopping amplitude $t$ near two valleys, leading to $H^{\rm honey}_{\rm Dir} (\vec{k})= v \; \left( i \gamma_0 \gamma_j k_j \right)$, where the momentum $|\vec{k}| \ll |{\bf K}|$ is measured from the respective valley, $v=\sqrt{3} t a/2$ is the Fermi velocity, and $\gamma_0=\tau_0 \beta_3$, $\gamma_1=\tau_3 \beta_2$, and $\gamma_2=\tau_0 \beta_1$ are mutually anticommuting four-component Hermitian matrices~\cite{graphenemicroscopic:1}. Two sets of Pauli matrices $\{ \beta_\mu \}$ and $\{ \tau_\mu \}$ operate on the sublattice and valley indices, respectively, with $\mu=0,1,2,3$ with $\beta_0$ and $\tau_0$ as two-dimensional identity matrices. To close the Clifford algebra of maximal five mutually anticommuting Hermitian matrices, we define two additional matrices $\gamma_3=\tau_1 \beta_2$ and $\gamma_5=\tau_2 \beta_2$ that together satisfy $\{ \gamma_\mu, \gamma_\nu \}= 2 \delta_{\mu \nu}$ for $\mu,\nu=0,1,2,3,5$.

The graphene-based Dirac theory enjoys a plethora of microscopic and emergent symmetries~\cite{graphenemicroscopic:2, graphenemicroscopic:3}. The Dirac Hamiltonian is invariant under the exchange of two sublattices ($S$) and valleys ($T$), generated by $\gamma_2$ and $\gamma_{51}$, respectively, under which $(k_x,k_y) \to (k_x,-k_y)$ and $(-k_x,k_y)$, where $\gamma_{\mu \nu}=i \gamma_\mu \gamma_\nu$. A rotational symmetry by $\pi/2$ about the $z$ axis is generated by $\gamma_{12}$, under which $(k_x,k_y) \to (-k_y, k_x)$. The time-reversal symmetry of the system is generated by ${\mathcal T}=\gamma_{51} {\mathcal K}$, where ${\mathcal K}$ is the complex conjugation, such that ${\mathcal T}^2=+1$ as it should be for spinless fermions. The system manifests a translational symmetry, which translates into a U(1) chiral symmetry in the continuum limit, generated by $\gamma_{35}$. Besides these microscopic symmetries, the planar Dirac Hamiltonian also enjoys an emergent SU(2) chiral symmetry, generated by $\{ \gamma_3, \gamma_5, \gamma_{35}\}$. The unitary particle-hole or sublattice symmetry of $H^{\rm honey}_{\rm Dir} (\vec{k})$ is generated by a set of anticommuting matrices $\{ \gamma_0, \gamma_{03}, \gamma_{05}, \gamma_{12} \}$, each of which thus represents a \emph{mass} matrix.

With mass matrices for Dirac fermions in graphene-based systems being identified, next we discuss their symmetry properties. Three mass matrices, forming the set $\{ \gamma_0, \gamma_{30}, \gamma_{50}\}$, transform as a vector under under SU(2) chiral rotations, while $\gamma_{12}$ is a scalar under the chiral transformation. The former mass orders preserve the ${\mathcal T}$ symmetry, while $\gamma_{12}$ is odd under ${\mathcal T}$. Physically, $\gamma_0$ represents a Semenoff mass resulting from a staggered pattern of fermionic density between the sites from two triangular sublattices of the honeycomb lattice~\cite{graphenemicroscopic:1}. Haldane's quantum anomalous Hall insulator mass, stemming from an intra-sublattice next-nearest-neighbor \emph{imaginary} currents, circulating in the \emph{opposite} directions on complementary sublattice is represented by $\gamma_{12}$~\cite{graphenemicroscopic:4}. Both of them preserves the translational symmetry, as $[\gamma_0, \gamma_{35}]=0=[\gamma_{12}, \gamma_{35}]$. On the other hand, the remaining chiral symmetry breaking masses $\{ \gamma_{30}, \gamma_{50}\}$ break the translational symmetry, correspond to two independent realizations of the commensurate modulations of the nearest-neighbor hopping amplitudes in the Kekul\'e pattern that enlarges the unit cell~\cite{graphenemicroscopic:5, graphenemicroscopic:6}.

\begin{figure}[t!]
\includegraphics[width=1.00\linewidth]{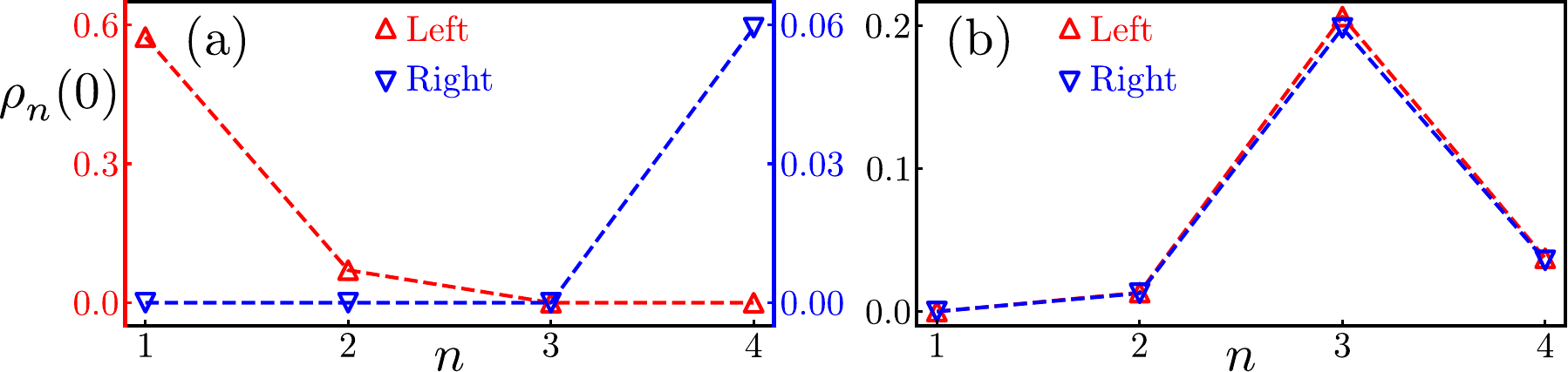}
\caption{Average local density of states per site $\rho_n(0)$ belonging to the $n$th generation of the $\{10,3 \}$ (Schl\"afli symbol) hyperbolic lattice (containing 2880 sites with a total of 4 generations) with open boundary condition (a candidate Dirac system on a negatively curved space) for the left (red triangles) and right (blue triangles) closest to zero-energy eigenmodes in the presence of (a) type-I ($\beta=1.0$) and (b) type-II ($\beta=0.5$) uniform non-Hermitian magnetic fields (NHMFs). The dashed lines are guides to eyes. The central decagon belongs to the first generation ($n=1$) and each successive layer of plaquettes constitutes its progressively next generation. For type-I NHMFs, on-site values of $\chi ({\vec{r}})= \beta (n-1)^2$, where $n$ plays the role of radial distance from the center of the system (Fig.~\ref{fig:1_modeldiagram}). For type-II NHMFs, the magnitude of the purely imaginary next-nearest-neighbor hopping, circulating in same direction on both sublattices (Fig.~\ref{fig:1_modeldiagram}), is determined by the smallest value of $n$ bordering each plaquette ($n_{\rm min}$) with $\chi(\vec{r})= \beta (n_{\rm min}-1)^2$. Results are qualitatively similar to the ones found in Euclidean Dirac systems (Fig.~\ref{fig:2_ldos}).
}~\label{fig:3_ldos_hyperbolic}
\end{figure}

Notice that on lattices, due to the fermion doubling~\cite{Fermiondoubling:1}, the minimal representation for ${\mathcal T}$-symmetric Dirac fermions in spinless systems is four-component and the number of mass operators is increased to four. Even though all the four mass matrices are bonafide representatives of $M$ in the construction of NHMFs, here we restrict to NH gauge fields that do not break the translational symmetry, thereby limiting the choices of $M$ to $\gamma_{12}$ and $\gamma_0$, respectively yielding its type-I and type-II realizations. Then in the presence of translational symmetry preserving NHMFs, the resulting Dirac operator takes the universal form
\begin{eqnarray}~\label{eq:NHMagLattice}
H^{\rm NH,\; honey}_{\rm Dir, \; type-a} (\chi, \vec{k}) &=& i \gamma_0 \gamma_j \left( k_j + M A_j \right) \nonumber \\
&=& e^{\chi(\vec{r}) \; N } \; H^{\rm honey}_{\rm Dir} (\vec{k}) \; e^{-\chi(\vec{r}) \; N},
\end{eqnarray}
where $M=\gamma_{12}$ and $N={\mathbf I}_4 \equiv \tau_0 \beta_0$ for $a={\rm I}$, and $M=\gamma_0$ and $N=-\gamma_{35}$ for $a={\rm II}$. Therefore, type-I NHMFs only foster normalizable left zero-energy eigenmodes near both the valleys, while the right zero-energy eigenmodes are always non-normalizable. By contrast, type-II NHMFs support normalizable right (left) zero-energy eigenmodes with Fourier components peaked near $+{\bf K}$ ($-{\bf K}$) valley, as $\gamma_{35} \equiv \tau_3 \beta_0$.

\emph{Numerical results}.~To anchor such continuum description based conclusions, we implement and diagonalize the final form of the NH operators shown in the second line of Eq.~\eqref{eq:NHMagLattice} on graphene's honeycomb lattice. The Dirac Hamiltonian $H^{\rm honey}_{\rm Dir} (\vec{k})$ results from the nearest-neighbor tight-binding model. We set $t=1$. To this end we choose the geometry of the honeycomb lattice shown in Fig.~\ref{fig:1_modeldiagram} with the zigzag (armchair) edges in the $x$ ($y$) direction along which periodic (open) boundary condition is imposed, such that there exist no edge modes in the pristine system that exclusively reside on the zigzag edges. We work with the Landau gauge with $\chi(\vec{r}) = \beta y^2$. The real parameter $\beta$ sets the strength of \emph{uniform} NHMFs. The $y=0$ line is chosen along the center of the system. For type-I NHMFs, the operator $\chi(\vec{r}) {\mathbf I}_4$ corresponds to a sublattice independent \emph{on-site} potential, whose spatial variation is shown in Fig.~\ref{fig:1_modeldiagram} for $\chi(\vec{r}) \propto y^2$. For type-II NHMFs, the operator $\chi(\vec{r}) \gamma_{35}$ stems from \emph{intrasublattice} imaginary currents between the next-nearest-neighbor sites, circulating in the \emph{same} directions on both sublattices, whose spatial variation for $\chi(\vec{r}) \propto y^2$ is shown in Fig.~\ref{fig:1_modeldiagram}. We numerically diagonalize the lattice-regularized NH operator associated with $H^{\rm NH,\; honey}_{\rm Dir, \; type-a} (\chi, \vec{k})$ from Eq.~\eqref{eq:NHMagLattice} for $a={\rm I}$ and II, and arrive at the following conclusions.

For type-I masslike NHMFs, we find that the closest to zero-energy ($\sim 10^{-11}$) left eigenmodes are sharply localized near the center of the system, see Fig~\ref{fig:2_ldos}(a), while their right eigenmodes reside at the $y$-directional boundaries of the system as shown in Fig~\ref{fig:2_ldos}(b). When the direction of the type-I NHMF is reversed with $\chi(\vec{r}) \to -\chi(\vec{r})$, the right (left) near zero-energy eigenmodes appear in the bulk (at the edges) of the system, which we do not show explicitly. Such findings are in agreement with our predictions from the continuum theory. On the other hand, for a type-II masslike NHMFs, the spatial profile of the left and right eigenmodes associated with the closest to zero-energy states ($\sim 10^{-7}$) are similar, as shown in Figs.~\ref{fig:2_ldos}(c) and~\ref{fig:2_ldos}(d), respectively, that are highly peaked near the boundary, but not on the edges of the system; compare with Figs.~\ref{fig:2_ldos}(b). This outcome can be justified in the following way. As a type-II NHMF supports right (left) normalizable/non-normalizable eigenmodes with Fourier components near $+/- {\bf K}$ ($-/+ {\bf K}$) valley, any numerical diagonalization yields a linear superposition of normalizable and non-normalizable right and left near zero-energy modes, which is localized away from the bulk and edges of the system, but peaked somewhat closer to the boundary. Overall, we find satisfactory agreements between the predictions on the zero-energy manifold for planar massless Dirac fermions coupled to masslike NHMFs and the corresponding lattice-based numerical findings. We also note that the number of near zero-energy modes ($N_0$) scales linearly with the number of plaquettes in the system ($N_{\rm plaq}$) in the presence of uniform type-I [Fig.~\ref{fig:2_ldos}(e)] and type-II [Fig.~\ref{fig:2_ldos}(f)] NHMFs with $N_{\rm plaq}$ being proportional to the flux enclosed.

\emph{Summary and Discussions}.~To summarize, from complementary continuum theory and lattice-based exact numerical diagonalization, here we show that gapless planar Dirac fermions coupled to masslike NHMFs accommodates a flat band of near zero-energy modes that features either self-normalizable right or left eigenmodes. On a lattice due to fermion doubling, one can realize a variety of such NHMFs. Specifically on graphene's honeycomb lattice we identify two realizations of such NHMFs that do not break the translational symmetry. Near each valley both of them support either left or right normalizable near zero-energy modes. However, due to the presence of valley degrees of freedom the masslike NHMF either supports normalizable left or right zero-energy eigenmodes near both valleys (type-I) or normalizable left and right zero-energy eigenmodes near complementary valleys (type-II). The degeneracy of the zero-energy manifold is determined by the flux of the NHMFs, enclosed by the system, bearing close connection to the celebrated ACIT~\cite{aharonovcasher:1}. Graphene-based Dirac systems features myriad mass orders~\cite{grapheneallmass:1, grapheneallmass:2}, each of which can give rise of NHMFs, including spin gauge fields. Similar to the situation with the ACIT for real/axial magnetic fields, which extends to the predicted $\{ 10,3\}$ (Schl\"afli symbol) hyperbolic Dirac system~\cite{aharonovcasher:3, aharonovcasher:4}, we arrive at similar conclusions for the spatial variation of near zero-energy modes with type-I and type-II NHMFs, see Fig.~\ref{fig:3_ldos_hyperbolic}.

Altogether, microscopic realizations of masslike NHMFs open a plethora of fascinating future avenues in theoretical physics that should be of interest in condensed matter and high-energy physics. We here outline only a few of those. For example, space and (imaginary) time dependent NH gauge fields would constitute the theoretical framework for NH quantum electrodynamics. Existence of a degenerate near zero-energy manifold supported by NHMF makes the system susceptible toward various orderings at weak coupling, a phenomenon that we name NH magnetic catalysis. As the vacuum expectation value of any ordering in NH systems is defined in terms of its biorthogonal expectation value, the zero-energy manifold with type-I NHMF should be robust against any spontaneous symmetry breaking, making it an ideal platform for the realization of strong interaction-driven symmetry preserving topological phases at both half and fractional fillings. By contrast, the manifold of normalizable zero-energy modes in the presence of type-II NHMFs \emph{exclusively} fosters a finite biorthogonal expectation value of valley-coherent translational symmetry breaking Kekul\'e mass order at weak coupling, in the spirit of NH magnetic catalysis, as then near $\pm {\bf K}$ valley system accommodates left and right normalizable zero modes, respectively. Our conclusions are consequential for three-dimensional Dirac or Weyl fermions coupled to $z$-directional NHMFs. In that case, the manifold of zero-energy modes becomes dispersive in the field direction, raising an intriguing possibility of NHMF induced chiral anomaly, when an electric field also applied in the same direction so that ${\bf E} \cdot {\bf B}$ is finite. Present work is thus the propelling fuel for the upcoming voyage through the vast unexplored rich relativistic landscape, immersed in masslike NH gauge fields.

Experimental verification of our predictions should be within the scope of presently available laboratory facilities. With the microscopic realizations of NHMFs being staged in Fig.~\ref{fig:1_modeldiagram} on the honeycomb lattice, resulting in spatially-modulated inter-sublattice hopping patterns with different amplitudes in the opposite directions (sourcing non-Hermiticity), it can be engineered on various highly tunable classical metamaterials, among which topolectric circuits and photonic lattices are most promising, as complex and non-reciprocal hopping, Peierls phase, etc.\ have already been realized therein~\cite{meta:1, meta:2, meta:3, meta:4}. Optical lattices of neutral atoms serve as yet another platform where our predictions in noninteracting systems can be tested, where various phenomena related to non-Hermiticity, complex hopping etc.\ have been emulated~\cite{opticalLattice:1, opticalLattice:2, opticalLattice:3}. Optical lattices are also ideal to showcase anticipated many-body phenomena within the zero-energy manifold with tunable interactions~\cite{opticalLattice:4, opticalLattice:5}.

\emph{Acknowledgments}.~This work was supported by NSF CAREER Grant No.\ DMR-2238679 of B.R. We thank Vladimir Juri\v{c}i\'c for critical comments on the manuscript.

\emph{Data availability}.~Numerical codes and data used and generated in this work are available in Ref.~\cite{chris:NHmagneticFields}.



\begin{thebibliography}{}

\bibitem{aharonovcasher:1} Y.\ Aharonov and A.\ Casher, Ground state of a spin-1/2 charged particle in a two-dimensional magnetic field, \href{https://doi.org/10.1103/PhysRevA.19.2461}{Phys.\ Rev.\ A {\bf 19}, 246 (1979)}.

\bibitem{aharonovcasher:2} B.\ Roy and I.\ F.\ Herbut, Inhomogeneous magnetic catalysis on graphene’s honeycomb lattice, \href{https://doi.org/10.1103/PhysRevB.83.195422}{Phys.\ Rev.\ B {\bf 83}, 195422 (2011)}.

\bibitem{aharonovcasher:3} B.\ Roy, Magnetic catalysis in weakly interacting hyperbolic Dirac materials, \href{https://doi.org/10.1103/PhysRevB.110.245117}{Phys.\ Rev.\ B {\bf 110}, 245117 (2024)}.


\bibitem{grapheneQHE:1} K.\ S.\ Novoselov, A.\ K.\ Geim, S.\ V.\ Morozov, D.\ Jiang, M.\ I.\ Katsnelson, I.\ V.\ Grigorieva, S.\ V.\ Dubonos, and A.\ A.\ Firsov, Two-dimensional gas of massless Dirac fermions in graphene, \href{https://www.nature.com/articles/nature04233}{Nature (London) {\bf 438}, 197 (2005)}.

\bibitem{grapheneQHE:2} Y.\ Zhang, Y.-W.\ Tan, H.\ L.\ Stormer, and P.\ Kim, Experimental observation of the quantum Hall effect and Berry's phase in graphene, \href{https://www.nature.com/articles/nature04235}{Nature (London) {\bf 438}, 201 (2005)}.


\bibitem{chiralanomalyTh:1} S.\ Adler, Axial-Vector Vertex in Spinor Electrodynamics, \href{https://journals.aps.org/pr/abstract/10.1103/PhysRev.177.2426}{Phys.\ Rev.\ {\bf 177}, 2426 (1969)}.

\bibitem{chiralanomalyTh:2} J.\ S.\ Bell and R.\ A.\ Jackiw, A PCAC puzzle: $\pi_0 \to \gamma \gamma$ in the
$\sigma$-model, \href{https://link.springer.com/article/10.1007/BF02823296}{Nuovo Cimento A {\bf 60}, 47 (1969)}.

\bibitem{chiralanomalyTh:3} H.\ B.\ Nielsen and M.\ Ninomiya, The Adler-Bell-Jackiw anomaly and Weyl fermions in a crystal, \href{https://www.sciencedirect.com/science/article/pii/0370269383915290?via%3Dihub}{Phys.\ Lett.\ B {\bf 130}, 389 (1983)}.


\bibitem{chiralanomalyExp:Review} For a review, see, N.\ P.\ Ong and S.\ Liang, Experimental signatures of the chiral anomaly in Dirac–Weyl semimetals, \href{https://www.nature.com/articles/s42254-021-00310-9}{Nat.\ Rev.\ Phys.\ {\bf 3}, 394 (2021)}.


\bibitem{axialMFTh:1} I.\ F.\ Herbut, Pseudomagnetic catalysis of the time-reversal symmetry breaking in graphene, \href{https://doi.org/10.1103/PhysRevB.78.205433}{Phys.\ Rev.\ B {\bf 78}, 205433 (2008)}.

\bibitem{axialMFTh:2} P.\ Ghaemi, J.\ Cayssol, D.\ N.\ Sheng, and A.\ Vishwanath, Fractional topological phases and broken time-reversal symmetry in strained graphene, \href{https://doi.org/10.1103/PhysRevLett.108.266801}{Phys.\ Rev.\ Lett.\ {\bf 108}, 266801 (2012)}.

\bibitem{axialMFTh:3} B.\ Roy and I.\ F.\ Herbut, Topological insulators in strained graphene at weak interaction, \href{https://doi.org/10.1103/PhysRevB.88.045425https://doi.org/10.1103/PhysRevB.88.045425}{Phys.\ Rev.\ B {\bf 88}, 045425 (2013)}.


\bibitem{axialMFEp:1} N.\ Levy, S.\ A.\ Burke, K.\ L.\ Meaker, M.\ Panlasigui, A.\ Zettl, F.\ Guinea, A.\ H.\ Castro Neto, and M.\ F.\ Crommie, Strain-Induced Pseudo–Magnetic Fields Greater Than 300 Tesla in Graphene Nanobubbles, \href{https://www.science.org/doi/10.1126/science.1191700}{Science {\bf 329}, 544 (2010)}.

\bibitem{axialMFEp:2} J.\ Lu, A.\ H.\ Castro Neto, and K.\ P.\ Loh, Transforming moir\'e blisters into geometric graphene nano-bubbles, \href{https://www.nature.com/articles/ncomms1818}{Nat.\ Commun.\ {\bf 3}, 823 (2012)}.

\bibitem{axialMFEp:3} K.\ K.\ Gomes, W.\ Mar, W.\ Ko, F.\ Guinea, and H.\ Manoharan, Designer Dirac fermions and topological phases in molecular graphene, \href{https://www.nature.com/articles/nature10941}{Nature (London) {\bf 483}, 306 (2012)}.

\bibitem{axialMFEp:4} M.\ Bellec, C.\ Poli, U.\ Kuhl, F.\ Mortessagne, and H.\ Schomerus, Observation of supersymmetric pseudo-Landau levels in strained microwave graphene, \href{https://www.nature.com/articles/s41377-020-00351-2}{Light Sci Appl {\bf 9}, 146 (2020)}.


\bibitem{Dirac:1} P.\ A.\ M.\ Dirac, The quantum theory of the electron, \href{https://doi.org/10.1098/rspa.1928.0023}{Proc.\ R.\ Soc.\ A {\bf 117}, 610 (1928)}.

\bibitem{Dirac:2} P.\ A.\ M.\ Dirac, A theory of electrons and protons, \href{https://doi.org/10.1098/rspa.1930.0013}{Proc.\ R.\ Soc.\ A {\bf 126}, 360 (1930)}.


\bibitem{NHDirac:1} V.\ Juri\v ci\' c and B.\ Roy, Yukawa-Lorentz symmetry in non-Hermitian Dirac materials, \href{https://doi.org/10.1038/s42005-024-01629-2}{Commun.\ Phys.\ {\bf 7}, 169 (2024)}.

\bibitem{NHDirac:2} S.\ A.\ Murshed and B.\ Roy, Quantum electrodynamics of non-Hermitian Dirac fermions, \href{https://doi.org/10.1007/JHEP01%282024%29143}{J.\ High Energy Phys.\ {\bf 01} (2024) 143}.


\bibitem{graphenemicroscopic:1} G.\ W.\ Semenoff, Condensed-Matter Simulation of a Three-Dimensional Anomaly, \href{https://journals.aps.org/prl/abstract/10.1103/PhysRevLett.53.2449}{Phys.\ Rev.\ Lett.\ {\bf 53}, 2449 (1984)}.

\bibitem{graphenemicroscopic:2} I.\ F.\ Herbut, Interactions and Phase Transitions on Graphene’s Honeycomb Lattice, \href{https://journals.aps.org/prl/abstract/10.1103/PhysRevLett.97.146401}{Phys.\ Rev.\ Lett.\ {\bf 97}, 146401 (2006)}.

\bibitem{graphenemicroscopic:3} I.\ F.\ Herbut, V.\ Juri\v{c}i\'c, and B.\ Roy, Theory of interacting
electrons on the honeycomb lattice, \href{https://journals.aps.org/prb/abstract/10.1103/PhysRevB.79.085116}{Phys.\ Rev.\ B {\bf 79}, 085116 (2009)}.  



\bibitem{graphenemicroscopic:4} F.\ D.\ M.\ Haldane, Model for a quantum Hall effect without Landau levels: Condensed-matter realization of the ``Parity Anomaly", \href{https://doi.org/10.1103/PhysRevLett.61.2015}{Phys.\ Rev.\ Lett.\ {\bf 61}, 2015 (1988)}.

\bibitem{graphenemicroscopic:5} C.-Y.\ Hou, C.\ Chamon, and C.\ Mudry, Electron fractionalization in two-dimensional graphenelike structures, \href{https://doi.org/10.1103/PhysRevLett.98.186809}{Phys.\ Rev.\ Lett.\ {\bf 98}, 186809 (2007)}.

\bibitem{graphenemicroscopic:6} B.\ Roy and I.\ F.\ Herbut, Unconventional superconductivity on honeycomb lattice: Theory of Kekule order parameter, \href{https://journals.aps.org/prb/abstract/10.1103/PhysRevB.82.035429}{Phys.\ Rev.\ B {\bf 82}, 035429 (2010)}.


\bibitem{Fermiondoubling:1} H.\ B.\ Nielsen and M.\ Ninomiya, Absence of neutrinos on a lattice: (I). Proof by homotopy theory, \href{https://www.sciencedirect.com/science/article/pii/0550321381903618}{Nucl.\ Phys.\ B {\bf 185}, 20 (1981)}.


\bibitem{grapheneallmass:1} S.\ Ryu, C.\ Mudry, C-Y.\ Hou, and C.\ Chamon, Masses in graphenelike two-dimensional electronic systems: Topological defects in order parameters and their fractional exchange statistics, \href{https://journals.aps.org/prb/abstract/10.1103/PhysRevB.80.205319}{Phys.\ Rev.\ B {\bf 80}, 205319 (2009)}.

\bibitem{grapheneallmass:2} A.\ L.\ Szab\'o and B.\ Roy, Extended Hubbard model in undoped and doped monolayer and bilayer graphene: Selection rules and organizing principle among competing orders, \href{https://journals.aps.org/prb/abstract/10.1103/PhysRevB.103.205135}{Phys.\ Rev.\ B {\bf 103}, 205135 (2021)}.


\bibitem{aharonovcasher:4} C.\ A.\ Leong and B.\ Roy,  Strained hyperbolic Dirac fermions: Zero modes, flat bands, and competing orders, \href{https://arxiv.org/abs/2511.16667}{arXiv:2511.16667}.


\bibitem{meta:1} R.\ El-Ganainy, K.\ G.\ Makris, M.\ Khajavikhan, Z.\ H.\ Musslimani, S.\ Rotter, and D.\ N.\ Christodoulides, Non-Hermitian physics and PT symmetry, \href{https://www.nature.com/articles/nphys4323}{Nat.\ Phys.\ {\bf 14}, 11 (2018)}.

\bibitem{meta:2} {\c{S}}.\ K.\ \"{O}zdemir, S.\ Rotter, F.\ Nori, and L.\ Yang, Parity–time symmetry and exceptional points in photonics, \href{https://www.nature.com/articles/s41563-019-0304-9}{Nat.\ Mater.\ {\bf 18}, 783 (2019)}.

\bibitem{meta:3} Y.\ Choi, C.\ Hahn, J.\ W.\ Yoon, and S.\ H.\ Song, Observation of an anti-PT-symmetric exceptional point and energy-difference conserving dynamics in electrical circuit resonators, \href{https://www.nature.com/articles/s41467-018-04690-y}{Nat. Commun. 9, 2182 (2018)}.

\bibitem{meta:4} T.\ Helbig, T.\ Hofmann, S.\ Imhof, M.\ Abdelghany, T.\ Kiessling, L.\ W.\ Molenkamp, C.\ H.\ Lee, A.\ Szameit, M.\ Greiter, and R.\ Thomale, Generalized bulk–boundary correspondence in non-Hermitian topolectrical circuits, \href{https://www.nature.com/articles/s41567-020-0922-9}{Nat.\ Phys.\ {\bf 16}, 747 (2020)}.


\bibitem{opticalLattice:1} M.\ Aidelsburger, M.\ Atala, M.\ Lohse, J.\ T.\ Barreiro, B.\ Paredes, and I.\ Bloch, Realization of the Hofstadter Hamiltonian with ultracold atoms in optical lattices, \href{https://doi.org/10.1103/PhysRevLett.111.185301}{Phys.\ Rev.\ Lett.\ {\bf 111}, 185301 (2013)}.

\bibitem{opticalLattice:2} M.\ Aidelsburger, Artificial gauge fields and topology with ultracold atoms in optical lattices, \href{https://iopscience.iop.org/article/10.1088/1361-6455/aac120}{J.\ Phys.\ B {\bf 51}, 193001 (2018)}.

\bibitem{opticalLattice:3} Q.\ Liang, D.\ Xie, Z.\ Dong, H.\ Li, H.\ Li, B.\ Gadway, W.\ Yi, and B.\ Yan, Dynamic Signatures of Non-Hermitian Skin Effect and Topology in Ultracold Atoms, \href{https://doi.org/10.1103/PhysRevLett.129.070401}{Phys.\ Rev.\ Lett.\ {\bf 129}, 070401 (2022)}.

\bibitem{opticalLattice:4} I.\ Bloch, J.\ Dalibard, and W.\ Zwerger, Many-body physics with ultracold gases, \href{https://journals.aps.org/rmp/abstract/10.1103/RevModPhys.80.885}{Rev.\ Mod.\ Phys.\ {\bf 80}, 885 (2008)}.

\bibitem{opticalLattice:5} T.\ Esslinger, Fermi-Hubbard physics with atoms in an optical lattice, \href{https://www.annualreviews.org/content/journals/10.1146/annurev-conmatphys-070909-104059}{Annu.\ Rev.\ Condens.\ Matter Phys.\ {\bf 1}, 129 (2010)}.




\bibitem{chris:NHmagneticFields} C.\ A.\ Leong, \href{https://github.com/CALeong/NH_Magnetic_Field}{Dirac fermions in non-Hermitian magnetic fields: Zero modes and index theorem}.

\end{thebibliography}
\end{document}